\documentclass[english,aps,pra,twocolumn,showpacs,superscriptaddress]{revtex4-2}

\usepackage{bm}
\usepackage{appendix}
\usepackage{graphicx}
\usepackage{amsmath}
\usepackage{amssymb}%
\usepackage{color}
\usepackage{lipsum}
\usepackage[colorlinks=true,citecolor=blue,urlcolor=blue]{hyperref}
\begin{document}
\title{Dynamically generating superflow in a bosonic ring via phase imprinting}
\date{\today}

\author{Ke-Ji Chen}
\email{chenkeji2010@gmail.com}
\affiliation{Zhejiang Key Laboratory of Quantum State Control and Optical Field Manipulation, Department of Physics, Zhejiang Sci-Tech University, Hangzhou 310018, China}

\author{Fan Wu}
\email{t21060@fzu.edu.cn}
\affiliation{Fujian Key Laboratory of Quantum Information and Quantum Optics, College of Physics and Information Engineering, Fuzhou University, Fuzhou, Fujian 350108, China}

\begin{abstract}
Phase imprinting enables the dynamic generation of superflow in bosonic atoms,  effectively overcoming traditional limitations such as vortex number constraints and heating effects. However, the mechanisms underlying superflow formation remain insufficiently understood.  In this work, we reveal these mechanisms by studying the time evolution of the transferred
total angular momentum and the quantized current throughout the phase imprinting process, achieved through
numerically solving the time-dependent Schr\"{o}dinger and Gross-Pitaevskii equations. We demonstrate that the Bose gas dynamically acquires angular momentum through the density depletion induced by the phase imprinting potential, whereas quantized currents emerge from azimuthal phase slips accompanied by complete density depletions. Regarding the impact of system parameters,  such as interactions,  we find that interactions hinder superflow formation, as the azimuthal density distribution becomes less susceptible to the phase imprinting potential. Our findings offer microscopic insights into the dynamic development of superflow during the phase imprinting process and provide valuable guidance for ongoing experimental efforts.
\end{abstract}

\maketitle

\section{Introduction}
Superflow (or persistent current) in ring geometries,  induced by vector potentials in superconductors~\cite{Tinkham-04}, or by synthetic gauge fields in neutral cold atoms~\cite{Dalibard-11, Spielman-14}, has significant application potentials due to the vanishing energy dissipation and the robustness of magnetic-flux quantization. For instance,  atomic superflow, realized in weakly interacting Bose-Einstein condensates (BECs) confined in toroidal traps,  can be useful for quantum sensing, offering an atomtronic analogue to the celebrated superconducting quantum interference devices (SQUIDs)~\cite{Phillips-11, Ryu-12, Campbell-14, Ryu-20, Kiehn-22}.  An annular or ring-shaped superfluid symbolizes the minimal realization of a matter-wave circuit, serving as a fundamental building block for future atomtronic devices~\cite{Amico-21, Amico-22}.  

The achievements of atomic superflows in BECs have spanned over a decade~\cite{Ryu-07, Hadzibabic-12, Hadzibabic-13} and generally rely on two schemes. One is rotation,  which has proven efficient in producing superflows with well-defined winding numbers~\cite{Wright-13},  but it is limited to preparing relatively small winding numbers of superflows and requires a long preparation time. The other is utilizing Laguerre-Gauss beams,  which carry angular momentum with helicoidal phases, to transfer angular momentum to the condensates through two-photon Raman processes~\cite{Hadzibabic-12, Andersen-06, Lin-18, Jiang-19}, and give rise to various intriguing phases~\cite{Pu-15, Sun-15, Qu-15, Chen-16, Hu-19, Chen-19, Pu-20, Duan-20, Chen-20, Wang-21, Chen-22, Han-22, Peng-22}. This scheme is efficient for 
the preparation of a specific superflow,  determined by the order of the angular momentum associated with the Laguerre-Gauss mode. However, the Raman process relies on the hyperfine structure of the atoms and introduces inevitable heating effects~\cite{Han-22, Cui-13},  which limits its application.

Phase imprinting, as a fundamental experimental technique,  has been proposed and utilized to induce superflow in bosonic atoms~\cite{Kumar-18}. More recently,  superflow in fermionic rings has also been demonstrated either by stirring strongly interacting Fermi gases~\cite{Cai-22} or through a phase imprinting process~\cite{Roati-22, Roati24}. In Ref.~\cite{Roati-22},  Del Pace \emph{et al}. introduce an angular-dependent phase imprinting potential and observe persistent current by measuring the winding number through an interferometric technique~\cite{Eckel-14, Beugnon-14, Mathew-15}. 
This work has attracted considerable interest in exploring the generation and stability of superflow in fermionic rings~\cite{Xhani-23, Chen-25, Xhani-25}.  
Building on the seminal experiment by Del Pace \emph{et al}.~\cite{Roati-22}, several fundamental questions naturally arise.  Can superflow emerge if we introduce such a phase imprinting potential into a Bose gas?  Whereas a conventional atomic gas is irrotational, superflow is distinguished by a finite circulation. How does the system evolve from a non-circulatory state to one with quantized circulation? Furthermore, how do system parameters influence the formation of superflow?

To address these questions and concentrate our discussions,  we consider a Bose gas confined by a ring-shaped potential and illuminated by a phase imprinting potential,  as realized in Ref.~\cite{Roati-22}.  We focus on the time evolution of the transferred total angular momentum and quantized current throughout the entire phase imprinting process. Many intriguing dynamical features, such as density depletions, phase slips of wave functions, and dynamical phase transitions, emerge.  We demonstrate that the density depletion,  induced by the phase imprinting potential, leads to an increase in the total angular momentum. Conversely, the quantized currents arise from azimuthal phase slips, accompanied by complete density depletions, driving dynamical phase transitions from a zero to a nonzero circulation state through the phase imprinting process.

This paper is organized as follows. Section~\ref{model} describes our system and analyzes its angular momentum dynamics.  Section~\ref{mechanism}  reveals the mechanism of superflow generation by analyzing the features of the wavefunction. In Section~\ref{impacts}, we investigate the impact of interactions and show that they hinder superflow formation. Finally, Section~\ref{conclusions} summarizes our main results.

\section{Model}
\label{model}
\begin{figure}[t]
\begin{center}
\includegraphics[width=0.46\textwidth]{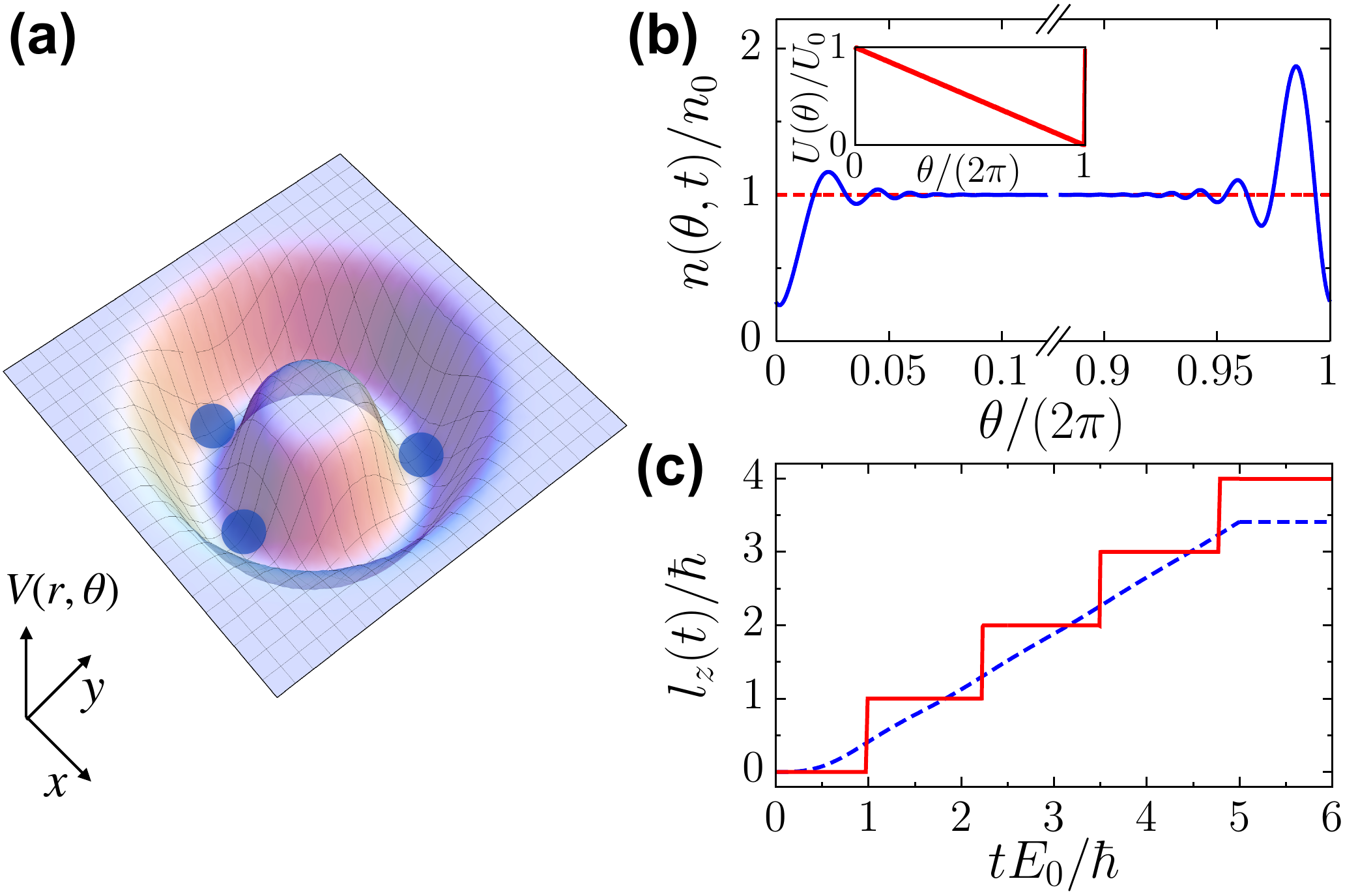}
\caption{(a) Schematic illustration of $V(r, \theta)$, which generates the ring-shaped potential. The blue dots represent the confined Bose atoms in the ring.  (b) Angular density profile $n(\theta,t)/n_0$  of a noninteracting Bose gas for $tE_0/
\hbar=0$ (red dashed curve) and $tE_0/\hbar=0.5$ (blue solid curve), respectilvely. The inset shows the profile of  $U(\theta)/U_0$ (red solid curve). (c) Time evolution of $l^{\rm tot}_z(t)/\hbar$ (blue dashed curve) and $l^{\varphi}_z(t)/\hbar$ (red solid curve). In (b)(c), we set $E_0=\hbar^2 /(M a^2)$ as an energy scale, where  $a$  serves as the characteristic length scale. Here, parameters are $U_0=5E_0, \Delta \theta=0.01\pi, R=10a, \tau E_0 /\hbar=5$, $n_0=1/(2\pi)$ and $n(\theta,t)$ satisfies $\int d\theta n(\theta,t)=1$.}
\label{Fig1}
\end{center}
\end{figure}

We consider a Bose gas with mass $M$  confined by an annular potential
  \begin{eqnarray}
V(r,\theta)  =  \sum_{j=1,2}V_0 \left (\tanh \left[\frac{(-1)^{j}(r-R_{j})}{d}\right]+1\right),
\label{V(r,theta)}
\end{eqnarray}
as illustrated in Fig.~\ref{Fig1}(a) in the $x-y$ plane,  and by $V(z)=M \omega^2_z z^2/2 $ in the $z$ direction. Here, $V_0$ denotes the trapping strength,  and  $R_{1}(R_2)$ represents the inner (outer) radius of  $V(r,\theta)$. The parameter $d$ is a length scale,  and $\omega_z$ is the trapping frequency along the $z$ axis. To capture the essential physics and simplify the calculation, we model the Bose gas confined in a ring-shaped potential. This configuration can be realized by ensuring $\hbar \omega_z$ exceeds any other energy scale, and by requiring $d \ll R_{j = 1, 2}$ with $R_1 \approx R_2$. Under these constraints, atomic degrees of freedom along the axial and radial directions are suppressed, resulting in a one-dimensional gas confined in a ring with a radius $R$, where $R \equiv (R_1 + R_2)/2$.

Phase imprinting is achieved via an angular-dependent potential $U(\theta)$,  as shown in the inset of Fig.~\ref{Fig1}(b), which is  given by
\begin{eqnarray}
U(\theta)=\left\{\begin{array}{l}  U_0 \Big [1-\frac{\theta}{2\pi -\Delta \theta}\Big] , \,\,  \theta \in [0,2\pi- \Delta \theta ],\\
\\
\frac{U_0}{\Delta \theta}\Big[\theta -(2\pi-\Delta \theta)\Big], \,\, \theta \in (2 \pi-\Delta \theta, 2\pi),  \end{array}\right.
\label{U_theta}
\end{eqnarray}
with $U_0$ the angular potential depth and  $\Delta \theta \ll 2\pi$.  As $[U(\theta), \hat{L}_z] \neq 0$ with $\hat{L}_z \equiv -i \hbar{\partial }/{\partial \theta}$,  $U(\theta)$  plays a crucial role in introducing angular momentum into the system.
Here,  we consider a specific scenario, in which   $U(\theta)$ is switched on (off) for $t \leqslant \tau $ ($t>\tau$),   with  $\tau$ the interval time.   Therefore, a  noninteracting Bose gas  can be described as
$ H_0(t) =   \int d\theta \psi^{\dag}(\theta,t){ \cal H}_s(\theta,t)\psi(\theta,t) $, 
where  $\psi(\theta,t)$ is the field operator,  satisfying the time-dependent Schr\"{o}dinger equation (TDSE)
\begin{eqnarray}
i \hbar \partial_t \psi(\theta,t) & = & {\cal H}_s(\theta,t) \psi(\theta,t), 
\label{time-Hs}
\end{eqnarray}
with 
\begin{align}
{\cal H}_s(\theta, t)= {\cal H}_s(\theta)+U(\theta) \vartheta (\tau-t),
\label{K_sigma}
\end{align}
where ${\cal H}_s(\theta) =  -\hbar^2/(2MR^2) \partial^2 /\partial \theta^2 $ and $\vartheta(x)$ is the Heaviside step function.  For the Bose gas, we choose the ground state of $H_0=  \int d\theta \psi^{\dag}(\theta) {\cal H}_s(\theta) \psi(\theta)$ as the initial state,  and we calculate the time evolution of $l^{\rm tot}_z(t)$,  where 
\begin{eqnarray}
l^{\rm tot}_{z}(t) \equiv \langle \psi(\theta,t) |\hat{L}_z |\psi(\theta,t)\rangle.
\label{lz_tot}
\end{eqnarray}
   
Before delving into the detailed calculation of the time evolution of $l^{\rm tot}_z(t)$,  it is helpful to analyze its equation of motion. According to Eqs.~(\ref{time-Hs}) and ~(\ref{lz_tot}),   
the equation of motion for $l^{\rm tot}_z(t)$ reduces to 
\begin{align}
\partial_t l^{\rm tot}_z(t)=U_0 \Big({\bar n}_{L}(t)-\bar{n}_{R}(t)\Big)\vartheta(\tau-t), 
\label{dLz_time}
\end{align}
where  
\begin{eqnarray}
\bar{n}_{L}(t) &=&  \frac{1}{2\pi-\Delta \theta} \int ^{2\pi-\Delta \theta}_0  d \theta  n(\theta,t), \\
\bar{n}_{R}(t) & =&  \frac{1}{\Delta \theta} \int^{2\pi}_{2\pi-\Delta \theta} d \theta n(\theta,t),
\end{eqnarray}
with $n(\theta,t)=|\psi(\theta,t)|^2$ representing the density profile. Here, $\bar{ n}_{L}(t)$ and $\bar{ n}_{R}(t)$ denote the average density for  $\theta \in [0,2\pi-\Delta \theta]$ and $\theta \in (2\pi-\Delta\theta, 2\pi)$, respectively.  As illustrated in Eq.~(\ref{dLz_time}), for $t \leqslant \tau$,  an inequality between $\bar{n}_{L}(t)$ and $\bar{n}_{R}(t)$ breaks the conservation of $l^{\rm tot}_z(t)$ and introduces angular momentum into the system. This imbalance naturally arises from the form of $U(\theta)$,  which inevitably introduces a density depletion for $\theta \approx 2\pi$ due to its sharp increase for $\theta \in (2\pi-\Delta \theta, 2\pi)$. 

We validate the above analysis by solving the TDSE. Specifically,  due to the ring geometry,  we expand $\psi(\theta,t)=\sum_{m}c_m(t)\Theta_m(\theta)$, where  $\Theta_m(\theta)=e^{i m \theta}/\sqrt{2\pi}$ with $m$ the magnetic qunatum number, and  $c_m(t)$ is the time-dependent coefficient.  This expansion transforms the TDSE into a matrix equation for $c_m(t)$,  given by
\begin{eqnarray}
i \hbar \partial_t c_m(t)  =  \sum_{m'} {\cal H}_{m,m'}(t)c_{m'}(t), 
\label{single-RK}
\end{eqnarray}
where  
$ {\cal H}_{m,m'}(t)={\cal K}_{m,m'}+f_{m,m'}\vartheta (\tau-t) $  with  ${\cal K}_{m,m'}=m^2 \hbar^2/(2MR^2)\delta_{m,m'}$ and 
\begin{align}
f_{m,m'} =\left\{\begin{array}{l} \frac{U_0}{2} , \,\,\,\,\, m=m',\\
\frac{U_0}{\Delta \theta (2\pi-\Delta \theta)}\frac{1-e^{i (m-m')\Delta \theta}}{(m-m')^2}, \,\,  m \neq m'.  \end{array}\right.
\label{fmmp}
\end{align}
Equation~(\ref{single-RK}) is then directly numerically solved using the fourth-order Runge-Kutta algorithm, and consequently $\psi(\theta,t)$ is determined as well.

Figure~\ref{Fig1}(b) shows the density profile $n(\theta,t)$, where a density depletion emerges for $\theta \approx 2\pi$ as expected.  Regarding the equation of motion for $l^{\rm tot}_z(t)$,  
as depicted in Fig.~\ref{Fig1}(c), we observe that $l^{\rm tot}_z(t)$  increases continuously from zero as our anticipation. The continuity of $l^{\rm tot}_z(t)$ 
can be attributed to the angular dependence of both the modulus and phase of $\psi(\theta, t)$.

The above analysis and calculations elucidate the origin of the injected angular momentum, yet the question of whether superflow can emerge remains ambiguous. To address this,  we define the angular momentum with respect to the phase of $\psi(\theta,t)$ as
\begin{eqnarray}
l^{\varphi}_z(t)  \equiv \frac{\hbar}{2\pi} \int d\theta \frac{\partial }{\partial \theta}\varphi(\theta,t),  
\label{lz_phi}
\end{eqnarray}
which is the quantized component of the current. Here, $\varphi(\theta,t)$  represents the phase of $\psi(\theta,t)$,  where $\psi(\theta,t)=|\psi(\theta,t)|e^{i \varphi(\theta,t)}$ with $|\psi(\theta,t)|$ the modulus of $\psi(\theta,t)$. The quantization of $l^{\varphi}_z(t)$ arises from  the single-valued nature of $\psi(\theta,t)$, which requires $\psi(0,t)=\psi(2\pi,t)$ since $\theta=0$ and $\theta=2\pi$ represent  the same point on the ring.  This leads to $\int d\theta \partial \varphi(\theta,t)/ \partial \theta=\varphi(2\pi,t)-\varphi(0,t)=2\pi \kappa$ with $\kappa$ the winding number.  In realistic calculation, $ \partial \varphi(\theta, t)/ \partial \theta$ in Eq.~(\ref{lz_phi}) can be extracted from the current density $j(\theta,t)$,  defined as
$ j(\theta,t)  =   [\psi^{\ast}(\theta,t) \partial \psi(\theta,t )/\partial \theta-h.c. ]/(2i) $
and $  \partial \varphi(\theta,t) /\partial \theta  =    j(\theta,t) /n(\theta,t)$.  Figure~\ref{Fig1}(c) shows the time evolution of $l^{\varphi}_z(t)$.  Three dynamical features of $l^{\varphi}_z(t)$ emerge. First,  $l^{\varphi}_z(t)$ becomes quantized as  expected.  Second,  jumps of $l^{\varphi}_z(t)$ occur during the phase imprinting process.  Third,  $l^{\varphi}_z(t)$ becomes stable at a long time, indicating that the persistent current state is robust. From these unique features, we conclude that superflow with different winding numbers emerge.

\section{Mechanism of superflow generation}
\label{mechanism}
\begin{figure}[b]
\begin{center}
\includegraphics[width=0.46\textwidth]{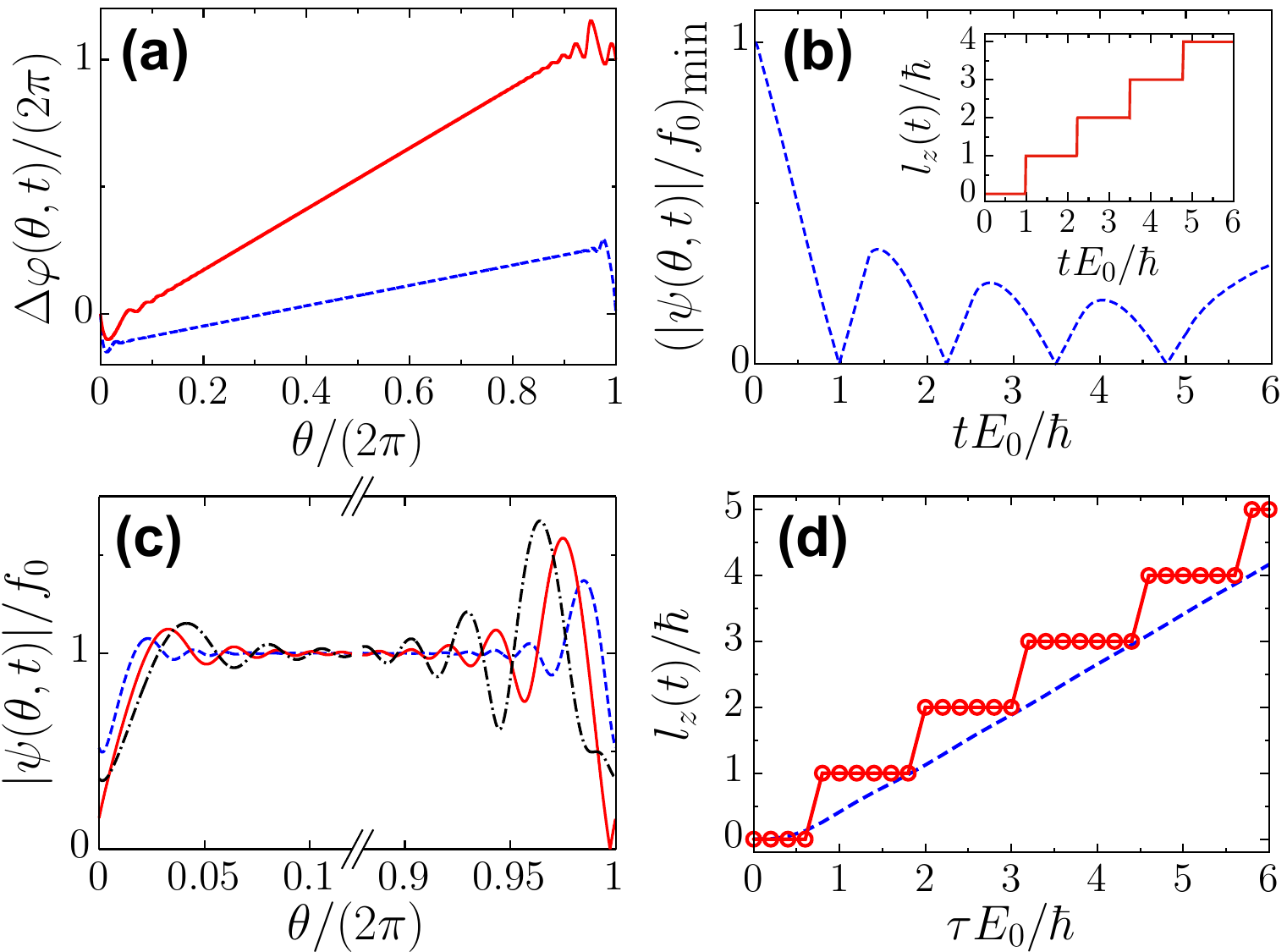}
\caption{(a) Time evolution profile of $\Delta \varphi(\theta,t)$ for  $tE_0/\hbar=0.5$ (blue dashed curve) and $t E_0 /\hbar=1.5$ (red solid curve) with a fixed quench time $\tau E_0/\hbar$. (b) Time evolution of $(|\psi(\theta,t)|/f_0)_{\text{min}}$ (blue dashed curve) with a fixed  $\tau E_0/\hbar$.  The inset shows the time evolution of $l^{\varphi}_z(t)$ (red solid curve). (c) Profile of $|\psi(\theta,t)|/f_0$ at $tE_0/\hbar=0.5$ (blue dashed curve), $tE_0/\hbar=0.98$ (red solid curve) and $tE_0/\hbar=1.5$ (black dash-dotted curve). (d) The injected angular momenta  $l^{\rm tot}_z(t)$ and $l^{\varphi}_z(t$)  at 
a long time. Here,  $f_0=1/\sqrt{2\pi}$; other parameters are the same as those in Fig.~\ref{Fig1}.}
\label{Fig2}
\end{center}
\end{figure}

To further elucidate the mechanism of superflow formation,  we calculate the phase  evolution of $\psi(\theta,t)$,  which is given by 
\begin{eqnarray}
\Delta \varphi(\theta,t)  =   \int^{\theta}_0\ d\theta \frac{\partial }{\partial \theta}\varphi(\theta,t).
\end{eqnarray}
As illustrated in Fig.~\ref{Fig2}(a), we show the time evolution of $\Delta \varphi(\theta,t)$ at different times as $\theta$ ranges from $0$ to $2\pi$. We observe that $\Delta \varphi(2\pi,t)=0$ for $tE_0/\hbar=0.5$;  however,  for $tE_0 /\hbar=1.5$,  $\Delta \varphi(2\pi,t)=2\pi$,  clearly indicating the occurrence of a phase slip between these two time points. 

We then consider the time evolution of $|\psi(\theta,t)|_{\rm min}$,  the minimal value of $|\psi(\theta,t)|$. As depicted in Fig.~\ref{Fig2}(b),  we observe that the jump time of $l^{\varphi}_z(t)$ [see the inset of Fig.~\ref{Fig2}(b)] and the time when $|\psi(\theta,t)|_{\rm min}=0$ are identical,  indicating that the phase slip occurs when the wave function (or density) becomes completely depleted.  We further confirm this observation by plotting the modulus profile of $\psi(\theta,t)$ at different times as illustrated in Fig.~\ref{Fig2}(c). We find that when $tE_0/\hbar=0.98$, a node point of $|\psi(\theta,t)|$ emerges at $\theta \approx 2\pi$,  with $|\psi(\theta, t)|$ being completely depleted, before or after this time, such as $tE_0/\hbar=0.5$ and $tE_0 /\hbar=1.5$,    no node point of $|\psi(\theta, t)|$ emerges.  Therefore,  the appearance of a node point of $|\psi(\theta,t)|$ can be considered as a criterion for the phase slip.  In Fig.~\ref{Fig2}(d), we present the relation between the stable $l^{\rm tot}_z(t)$  ($l^{\varphi}_z(t))$ and $\tau$ at long times,  meaning $t \gg \tau$.  Figure~\ref{Fig2}(d) indicates that  a specific quantized current state can be obtained by {\color{red}tuning} the duration  $\tau$.

From the detailed analysis above, we conclude that the generation of superflow arises from phase slips,  accompanied by the depletions of the wave functions, driving phase transitions from a zero-circulation state to a state with nonzero circulation during the phase imprinting process. By tuning the duration $\tau$, quantized current states with different phase windings can be obtained.


\section{Impacts of interactions}
\label{impacts}
Regarding realistic systems,  interactions are inevitable. Here, we consider an $s$ wave interaction, and    the full Hamiltonian is given by
$H(t) =  H_{0}(t)+H_{\rm int}(t)$, 
where 
\begin{eqnarray}
H_{\rm int}(t)=\frac{g}{2} \int d\theta \psi^{\dag}(\theta,t)\psi^{\dag}(\theta,t)\psi(\theta,t)\psi(\theta,t) 
\end{eqnarray}
with $g$ the interaction strength. Under the mean field approximation,  the Hamiltonian reduces to 
\begin{eqnarray}
H_{\rm MF}(t) =  \int d\theta  \phi^{\ast}_{0}(\theta,t) \Big[ {\cal H}_s(\theta,t)+\frac{g}{2}n_0(\theta,t) \Big]\phi_{0}(\theta,t), 
\nonumber\\
\label{HmF}
\end{eqnarray}
where $n_0(\theta,t)=|\phi_0(\theta,t)|^2$ is the condensate density of the interacting Bose gases with $\phi_0(\theta,t)$ the condensate wave function.   Here, $\phi_0(\theta,t)$ satisfies the  time-dependent Gross-Pitaevskii equation,  which is given by
\begin{eqnarray}
i \hbar \partial_t \phi_0(\theta,t)  =  \Big[ {\cal H}_s(\theta,t)+gn_0(\theta,t)\Big]  \phi_0(\theta,t).
\label{GP}
\end{eqnarray}

\begin{figure}[b]
\begin{center}
\includegraphics[width=0.45\textwidth]{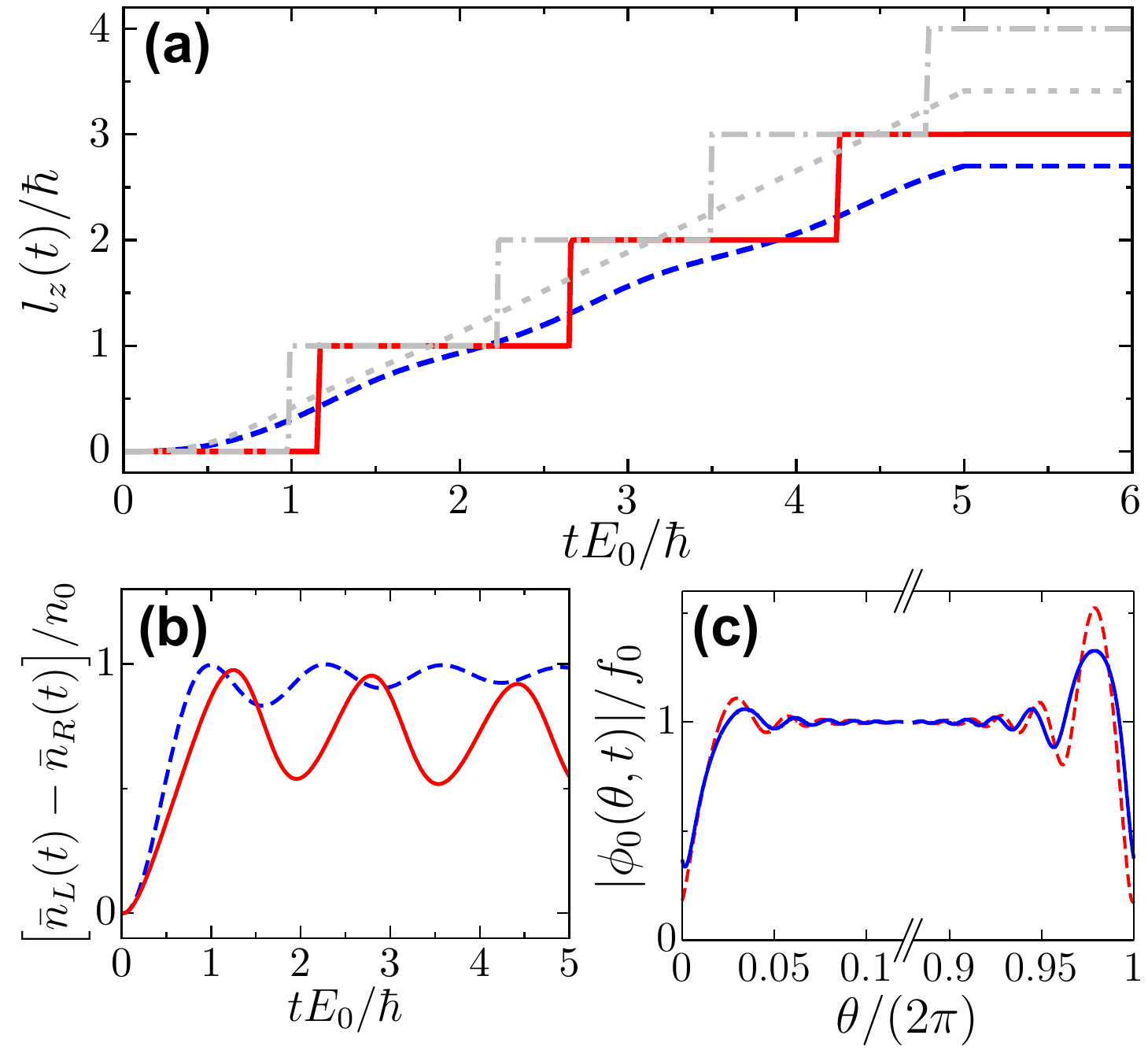}
\caption{(a) Time evolution of  $l^{\rm tot}_z(t)/\hbar$ and  $l^{\varphi}_z(t)/\hbar$ under different interaction strengths with a fixed $\tau E_0/\hbar$. The blue dashed (gray dashed) curve and red solid (gray dash-dotted) curve denote $l^{\rm tot}_z(t)/\hbar$ and $l^{\varphi}_z(t)/\hbar $,   respectively,   for  $g=10E_0$ ($g=0$).  (b) Time evolution of $[\bar{n}_L(t)-\bar{n}_R(t)]/n_0$ under different interaction strengths,  the red solid (blue dashed) curve denotes the time evolution of $[\bar{n}_L(t)-\bar{n}_R(t)]/n_0$ for  $g=10E_0$ ($g=0$).  (c) Density profile under different interaction strengths at the same time. The blue solid (red dashed) curve denotes the profile of $|\phi_0(\theta,t)|/f_0$ for $g=10E_0$ ($g=0$) at $tE_0/\hbar=0.8$. Here,  other parameters are the same as those in Fig.~\ref{Fig1}.}
\label{Fig3}
\end{center}
\end{figure}

Similar to solving the TDSE, we expand $\phi_0(\theta,t)=\sum_m c_m(t)\Theta_m(\theta)$ and substitute this expansion  into Eq.~(\ref{GP}), yielding 
\begin{eqnarray}
i \hbar \partial_t c_m(t)  =  \sum_{m'}\Big[{\cal H}_{m,m'}(t)+d_{m,m'}(t)\Big]c_{m'}(t), 
\label{cmt}
\end{eqnarray}
where 
\begin{eqnarray}
d_{m,m'}(t)  =  \frac{g}{2\pi} \int d\theta n_0(\theta,t)e^{i (m'-m)\theta}.
\end{eqnarray}
Equation~(\ref{cmt}) can be solved using the fourth-order  Runge-Kutta method, allowing us to  obtain $\phi_0(\theta,t)$ as well.

Analogous to the noninteraction case, as illustrated in Eqs.~(\ref{lz_tot}) and~(\ref{lz_phi}), we define the total angular momentum $l^{\rm tot}_z(t)$ and the quantized current $l^{\varphi}_z(t)$ in the interaction case by replacing the field operator $\psi(\theta, t)$ with the condensate wave function $\phi_0(\theta, t)$. Here, we choose the ground state of an interacting Bose gas without phase imprinting potential as the initial state, and calculate the time evolution of $l^{\rm tot}_z(t)$ and $l^{\varphi}_z(t)$.  As shown in Fig.~\ref{Fig3}(a),  by comparing the results with and without interactions, we find that the presence of interactions suppress both $l^{\rm tot}_z(t)$ and $l^{\varphi}_z(t)$. This can be understood as follows: interactions favor a homogeneous density, which reduces the density depletion and,  in turn,  suppresses the increase of $l^{\rm tot}_z(t)$. This understanding is further confirmed by numerical calculating the density difference between $\bar{n}_L(t)$ and  $\bar{n}_R(t)$, as illustrated in Fig.~\ref{Fig3}(b). Here,  we present the time evolution of $\bar{n}_L(t) - \bar{n}_R(t)$ under different interaction strengths and demonstrate that the stronger interaction strength leads to a smaller difference between  $\bar{n}_L(t)$ and $\bar{n}_R(t)$. According to Eq.~(\ref{dLz_time}), this reduced difference results in a slower increase in $l^{\rm tot}_z(t)$. Ultimately, the smaller $l^{\rm tot}_z(t)$  gives rise to  a diminished $l^{\varphi}_z(t)$. Figure~\ref{Fig3}(c)  presents  $|\phi_0(\theta, t)|/f_0$ under different interaction strengths at the same time. When contrasted with the results in the absence of interactions, we observe that the interactions reduce the oscillation of the wave function, which confirms our analysis that interactions favor a homogeneous density, making the system less susceptible to the phase imprinting potential,  and ultimately suppressing both $l^{\rm tot}_z(t)$ and $l^{\varphi}_z(t)$.


\section{Conclusions}
\label{conclusions}
We investigate the generation of superflow and dynamic transitions induced by angular phase imprinting in a Bose gas. We demonstrate that dynamic transitions occur between the zero-circulation  state and finite-circulation current states through the phase imprinting technique. Our microscopic approach reveals that the transferred angular momentum arises from the density depletion,  induced by the phase imprinting potential,  whereas transitions between states with different quantized persistent currents are driven by phase slips and wave function (or density) depletions in the angular direction. We further investigate the impact of interaction strength in the phase imprinting process and find interactions suppress the formation of superflow. Our results provide a microscopic understanding for the dynamic generation of superflow, several distinctive features that can be probed through time-of-flight imaging combined with interferometric techniques. These insights may contribute to refining current experimental protocols for superflow preparation and control.

\section*{acknowlegement}
We acknowledge fruitful discussions with professor Wei Yi. This work is supported by the Natural Science Foundation of China (Grants No. 12104406 and No. 12204105). 
K.C. acknowledges support by Zhejiang Provincial Natural Science Foundation (Grant No. ZCLMS25A0401) and the startup grant of Zhejiang Sci-Tech University (Grant No. 21062338-Y).  F. W. is supported by the Natural Science Foundation of Fujian Province (Grant No. 2022J05116).

\bibliographystyle{iopart-num.bst} 
\bibliography{myref}

\end{document}